\documentclass[twocolumn]{aastex631}

\begin{document}

\title{Changing-Look AGN Behaviour Induced by Disk-Captured Tidal Disruption Events}

\author{Yihan Wang}
\affiliation{Nevada Center for Astrophysics, University of Nevada, Las Vegas, NV 89154}
\affiliation{Department of Physics and Astronomy, University of Nevada Las Vegas, Las Vegas, NV 89154, USA}

\author{Douglas N. C. Lin}
\affiliation{Department of Astronomy \& Astrophysics, University of California, Santa Cruz, CA 95064, USA}

\author{Bing Zhang}
\affiliation{Nevada Center for Astrophysics, University of Nevada, Las Vegas, NV 89154}
\affiliation{Department of Physics and Astronomy, University of Nevada Las Vegas, Las Vegas, NV 89154, USA}

\author{Zhaohuan Zhu}
\affiliation{Nevada Center for Astrophysics, University of Nevada, Las Vegas, NV 89154}
\affiliation{Department of Physics and Astronomy, University of Nevada Las Vegas, Las Vegas, NV 89154, USA}



\begin{abstract}
Recent observations of changing-look active galactic nuclei (AGN) hint at a frequency of accretion activity not fully explained by tidal disruption events (TDEs) stemming from relaxation processes in nucleus star clusters (NSCs), traditionally estimated to occur at rates of $10^{-4}$ to $10^{-5}$ yr$^{-1}$ per galaxy. In this letter, we propose an enhanced TDE rate through the AGN disk capture process, presenting a viable explanation for the frequent transitions observed in changing-look AGN. Specifically, we investigate the interaction between the accretion disk and retrograde stars within NSCs, resulting in the rapid occurrence of TDEs within a condensed time frame. Through detailed calculations, we derive the time-dependent TDE rates for both relaxation-induced TDE and disk-captured TDE. Our analysis reveals that TDEs triggered by the disk capture process can notably amplify the TDE rate by several orders of magnitude during the AGN phase. This mechanism offers a potential explanation for the enhanced high-energy variability characteristic of changing-look AGNs.

\end{abstract}



\section{Introduction} \label{sec:intro}
Active Galactic Nuclei (AGNs) are distinctly categorized into Type 1 and Type 2 classes based on their different emission-line characteristics \citep{Seyfert1943,Khachikian1971}. The distinguishing feature of Type 1 AGNs is the presence of both broad and narrow emission lines, whereas Type 2 AGNs exhibit only narrow emission lines \citep{Netzer2015}.

Numerous AGNs have been documented to experience transitions across various spectral types, a phenomenon classified as changing-look (CL) AGNs \citep{Maciejewski2004,Stern2018,Sheng2020}. This phenomenon poses significant challenges to the generally accepted orientation-centric AGN unified model \citep{Antonucci1993, Urry1995}, where the central engine in Type 2 AGNs is obscured by a dusty torus situated along the observer's line of sight. Moreover, this phenomenon questions the conventional disk model, particularly casting doubts on its explanations concerning the disk viscosity \citep{Lawrence2018}.

The physical origin of CL-AGNs is still under debate. The prominent theories have been proposed include: (1) fluctuating obscuration influenced by the movements of obscuring materials, possibly due to the dusty toroidal structure with a patchy distribution obscuring the BLR \citep{Nenkova2008,Elitzur2012} or accelerating outflows \citep{Shapovalova2010}; (2) changes in accretion rates, according to which an AGN undergoes a series of evolutionary phases \citep{Penston1984,Elitzur2014,Yang2018,Wang2018}; (3) TDEs where a star is disrupted by the supermassive black hole (SMBH) \citep{Eracleous1995,Merloni2015,Blanchard2017}.

In the study of changing-look (CL) AGNs, polarization measurements provide critical insight into the mechanisms driving the type transitions. \citet{Hutsemekers2017} and \citet{Marin2017}  argued that high linear polarization would be observed if the transitions were triggered by obscuration. However, polarization studies on CL-AGNs \citep{Hutsemekers2017} revealed no significant polarization, indicating that variable obscuration was not the cause behind the type transition. This aligns with recent observations of a swift "turn-on" of the quasar J1554+3629 documented by the intermediate Palomar Transient Factory (iPTF), which noted a tenfold increase in UV and X-ray continuum flux over a period of less than a year \citep{Gezari2017}. This rapid alteration suggests an intrinsic shift in the accretion rate rather than external obscuration factors.

Further supporting this theory, systematic reviews of data from the Sloan Digital Sky Survey (SDSS) uncovered additional CL quasars exhibiting similar traits \citep{MacLeod2016, Ruan2016}. The changes in the accretion rate appeared to more adequately explain the observed variations in transition time scales and emission-line properties compared to fluctuating dust obscuration. 

Recent evolutionary CL-AGN studies have shown that TDEs predominantly occur in E+A galaxies \citep{Dodd2021}. The presence of significant Balmer absorption in these galaxies indicates a considerable starburst population aged approximately 0.1-1 Gyr. Despite constituting about 2$\%$ of the local galaxy population, these E+A galaxies are host to over half of the detected TDE candidates identified up to date \citep{French2020,Hammerstein2021}. This pattern indicates that a dynamic mechanism is at play, potentially amplifying the TDE rate efficiently within 0.1-1 Gyr \citep{Law-Smith2017}. Notably, a small but significant fraction of E+A TDE hosts exhibit pre-flare AGN activity, as claimed in recent research \citep{French2020}, hinting that the occurrence of TDEs might be boosted by the existence of an earlier accretion disk. As a result, the potential of TDEs in driving CL phenomena, as proposed by various researchers, holds considerable promise. 

In this letter, we propose a new mechanism that occurs during the AGN phase, where the pre-existing AGN disk interacts with stars in the NSC, leading to rapidly captured TDEs. This mechanism can potentially explain the enhanced TDE rate required during the AGN phase in CL-AGNs. The letter is structured as follows: In Section 2, we discuss the TDE rate arising from NSC relaxations from dormant SMBHs. Following that, in Section 3 we investigate the time-dependent TDE rate from disk TDE and captured TDE during the AGN phase. Utilizing the calculated TDE rate, we estimate the fraction of CL-AGNs in the overall AGN population in Section 4. Finally, we discuss and summarize our findings in Section 5.

\section{TDE rate in quiescent galaxies}
TDEs are crucial for the growth of massive black holes in stellar systems, especially in nuclei clusters. These events occur when stars come within a distance smaller than the tidal radius, $r_t$, which is given by
\begin{eqnarray}
r_t \sim \left(\frac{M_\bullet}{M_*}\right)^{1/3}R_*\,,
\end{eqnarray}
where, $M_\bullet$ is the black hole mass, and  $M_*$ and $R_*$ represent the mass and radius of the disrupted stars, respectively. 
The maximum specific angular momentum allowed for tidal disruption is  $L_{\rm TDE}\sim \sqrt{2GM_{\bullet}r_t}$.

In galaxies with dormant SMBHs, relaxation processes in nuclear star clusters (NSCs), such as two-body and resonance relaxations, primarily fuel the TDEs. The NSC surrounding an SMBH follows a power-law density profile \citep{Merritt2013}:
\begin{eqnarray}
n &=& \frac{(3-\gamma_{\rm NSC})}{2\pi}\frac{M_{\bullet}}{m}r_{\rm m}^{-3}\left(\frac{r}{r_{\rm m}}\right)^{-\gamma_{\rm NSC}}\,,
\label{eq:ndistri} \\
r_{\rm m} &=& \frac{GM_{\bullet}}{\sigma_{\rm NSC}^2}\,,
\label{eq:rm}
\end{eqnarray}
where $r$ is the distance from the SMBH,
$\gamma_{\rm NSC}$ is the radial density power-law index, $m$ is the average mass of the stars in NSC chosen to be $1M_\odot$ and $\sigma_{\rm NSC}$ is the velocity dispersion of the NSC based on the empirical $M_\bullet$-$ \sigma_{\rm NSC}$ relation \citep{Kormendy2013}. 

In the NSC, the two-body relaxation process causes energy and angular momentum fluctuations in stars, characterized by the relaxation time $t_{\rm rel}$, defined as \citep{Jeans1913,Jeans1916,Chandrasekhar1942,Benney1987}

\begin{eqnarray}
&&\frac{\Delta E}{E}\sim \sqrt{\frac{t}{t_{\rm rel}}}\,,\\
&&\frac{\Delta L}{L_{\rm max}(E)}\sim \sqrt{\frac{t}{t_{\rm rel}}}\,,\\
&&t_{\rm rel}\sim\frac{M^2_{\bullet}}{m^2N\ln{\Lambda}}T\,,
\end{eqnarray}
where $E=-\frac{G(M_\bullet+m)}{2a}$ is the specific binding energy with orbit semi-major axis $a$. $L_{\rm max}=\sqrt{G(M_\bullet+m)a}$ is the maximum  specific angular momentum for a given $E$, $N={2M_{\bullet}}/{m}$ is the number of stars within the NSC, $\Lambda$ is the Coulomb logarithm, and $T$ is the orbital period. For circular orbits, $a$ is equivalent to $r$ in Eq~\ref{eq:ndistri}.  

The TDE rate at a specific energy $E$ can be defined using $\lambda(E)$, which represents the fraction of stars with that particular energy $E$ that undergo disruption per orbital period \citep{Lightman1977,Frank1976,Rees1988}:
\begin{eqnarray}
\dot{N}_{\rm TDE}\sim\int_{S_{\rm TDE}}\lambda (E)\frac{n(E)}{T}\frac{dE}{ dr^\prime}r^{\prime2}dr^\prime d\Omega\,.
\label{eq:relaxlossrate}
\end{eqnarray}
Here, $S_{\rm TDE}$ represents the regions of the NSC that contribute to the TDE, $\Omega$ is the solid angle, and $n(E)$ is the number density of energy states.

The TDE rate depends on whether the loss cone is `full' or `empty'. The full loss cone regime occurs when $\Delta L$ is significantly larger than $L_{\rm TDE}$, leading to a faster relaxation refilling rate than TDE consumption. In contrast, the empty loss cone regime happens when $\Delta L$ is much smaller than $L_{\rm TDE}$, resulting in a faster SMBH star consumption rate compared to the relaxation refilling rate.

The inner region of the NSC is in the empty loss cone regime due to a smaller semi-major axis and $\Delta L$, whereas the outer region is in the full loss cone regime, where $\Delta L$ typically exceeds $L_{\rm TDE}$.
In the empty and full loss cone regimes \citep{Lightman1977,Frank1976,Rees1988,Magorrian1999,Stone2016}:
\begin{eqnarray}
\lambda_{\rm empt}(E)&\sim& \frac{\Delta L^2}{L^2_{\rm max}(E)\ln\left(\frac{L_{\rm max}(E)}{L_{\rm TDE}}\right)}\,,\\
\lambda_{\rm full}(E)&\sim& \frac{L^2_{\rm TDE}}{L^2_{\rm max}(E)}\,.
\end{eqnarray}

In the early stage when the inner loss cones have not been completely depleted, the overall TDE rate of the NSC is dominated by the empty loss cone region as $\lambda_{\rm empt}(E)\gg \lambda_{\rm full}(E)$. As the inner loss cones are completely depleted, the overall TDE rate enters into a steady state, where the TDEs are supplied relaxations in the full loss cone regime.  The steady-state TDE rate from two-body relaxation can be approximated by substituting $\lambda (E)= \lambda_{\rm full} (E)$ in Eq (\ref{eq:relaxlossrate}).  The contributed region of the NSC is $r\in[r_c, r_m]$, $\Omega=4\pi$, where $r_c$ is the critical radius 
that $\Delta L=L_{\rm TDE}$ and $r_m$ (Eq. \ref{eq:rm}) is the size of the NSC. 

\begin{figure}
    \includegraphics[width=\columnwidth]{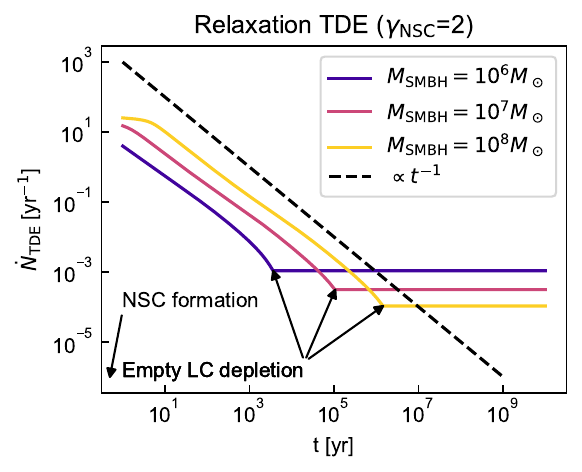}\\
    \caption{TDE rate from relaxations around dormant SMBHs as a function of time. The radial power-law index of the NSC $\gamma_{\rm NSC}=2$. }
    \label{fig:relx}
\end{figure}
Figure~\ref{fig:relx} shows the TDE rate from relaxations as a function of time for different SMBH masses and NSC profiles. As indicated in this figure, the TDE rate can be higher than the steady-state TDE rate in the early stage (with $\lambda(E)=\lambda_{\rm empty} (E)$) before the inner loss cone depletion.

\section{Enhanced TDE rate in AGN phase}
Once the AGN activates, an accretion disk forms around the SMBH, significantly altering the orbital evolution of stars in the NSC and affecting the TDE rates during the AGN phase.
\subsection{Disk model}
The accretion disk can be simply characterized by three parameters: the accretion rate efficiency $\lambda_{\rm d}$, the Toomre parameter $Q_{\rm d}$, and the viscosity parameter $\alpha_{\rm d}$, which represents the efficiency of angular momentum transport from disk turbulence. The accretion rate can be approximated as follows:
\begin{eqnarray}
\dot{M}&=&\lambda_{\rm d} \dot{M}_{\rm Edd} = 2.2\lambda_{\rm d}\left(\frac{M_{\bullet}}{10^8M_\odot}\right) M_\odot/{\rm yr}\,,\\
\dot{M}_{\rm Edd} &=& \frac{L_{\rm Edd}}{\epsilon_{\rm d} c^2}\,,
\end{eqnarray}
where, $\dot{M}$ represents the accretion rate, $\dot{M}_{\rm Edd}$ denotes the Eddington accretion rate, $M_{\rm SMHB}$ represents the mass of the SMBH, $L_{\rm Edd}$ is the Eddington luminosity, and $\epsilon_{\rm d}$ is a constant assumed to be 0.1.  

The disk surface density $\Sigma$ can be described by \citep{Papaloizou1995},
\begin{eqnarray}
\Sigma &=& \frac{\dot{M}}{2\pi r v_r}\,,\\
v_r &=& \alpha h^2 \sqrt{\frac{GM_{\bullet}}{r}}\,,\\
h &=& H / r = \left(\frac{Q_{\rm d}\dot{M}}{2\alpha_{\rm d}M_{\bullet}\Omega_{\rm d}}\right)^{1/3}\,,
\end{eqnarray}
where, $r$ represents the distance from the SMBH, $H$ is the scale height, and $\Omega_{\rm d}$ denotes the orbital frequency given by $\sqrt{GM_{\bullet}/r^3}$. The parameters $\lambda_{\rm d}$, $\alpha_{\rm d}$, and $Q_{\rm d}$ are assumed to be constant values set 
to 1 \citep{lin1987}, although the $\alpha_{\rm d}$ value is quite uncertain. The $\alpha_{\rm d}$ value from magneto-rotational instability depends on the net field strength\citep{Zhu2018}, while the $\alpha_{\rm d}$ value from gravitational instability depends on the disk cooling rate \citep{Gammie2001, deng2020}.

\subsection{Disk TDE from in-situ orbits}
During the AGN phase, a portion of stars in the NSC with orbit inclination below the AGN disk specific scale height $h$ intersect with the AGN disk \citep{artymowicz1993}. Those initially fully embedded stars undergo continuous interactions with the disk.

For prograde stars, they evolve into circular orbits due to aerodynamic drag and dynamical friction, eventually 
co-rotating with the disk. Once the star becomes corotating with the disk, both aerodynamic drag and dynamical friction become very weak. However, the Lindblad resonance from the density wave inspired by the star comes into play, slowly damps the orbital semi-major axis 
on the timescale of  \citep{tanaka2002}
\begin{eqnarray}
\tau_{\rm pro}\sim h^2\frac{M_{\bullet}}{\Sigma \pi r^2}\frac{M_{\bullet}}{m}T  \simeq {h Q_{\rm d}} {M_\bullet\over m} T.
\end{eqnarray}
 In a turbulent disk, the orbital decay timescale may be lengthened by stochastic 
migration \citep{nelson2005, baruteau2010, Zhu2013}.
Due to the orbit circularization and the long timescale of semi-major axis damping from the Lindblad resonance, solar-type stars
($m \simeq 1 M_\odot$) with prograde 
orbits have a negligible contribution to the TDE rate.  However, self-regulated star formation \citep{goodman2003, 
chen2023} can significantly increase the mass budget of the embedded prograde stars.  Moreover, stars
formed or captured by the disk
also gain mass through disk-gas accretion \citep{cantiello2021,li2021} which would reduce $\tau_{\rm pro}$.
But these stars evolve off the main sequence within $\tau_\star \sim 3-5$ Myr and return most of its envelope to the disk
\citep{alidib2023} before they migrate over significant radial range from their birth place. Indeed, the observed broad-line ratios 
indicate super solar metallicity, independent of redshifts and the observed differential metallicity between the broad-line
and narrow-line regions suggests {\it in situ} pollution in AGN disk \citep{huang2023}. These data provide
support for the assumption that the stars formed in AGN disks generally demise before they venture to and 
undergo tidal disruption in the proximity of SMBHs so that $\lambda_{\rm pro}$ is negligible.

Retrograde stars, moving against the disk rotation, experience significant aerodynamic drag, rapidly decreasing their semi-major axis. 
With a geometric cross section $\pi R_\star^2$ where $R_\star$ is the stellar radius, this fast-damping timescale is estimated as 
\begin{equation}
\tau_{\rm ret}\sim \frac{\Sigma_*}{\Sigma}hT
 \sim \left( { m \over M_\bullet} \right) ^{1/3}  {r^2 \over r_{\rm t}^2}
Q_{\rm d} T \sim {m \over M_\bullet} {R_{\rm R} \over H} {r^2 \over r_{\rm t} ^2} \tau_{\rm _{\rm pro}}\,,
\end{equation}
where $\Sigma_*=\frac{m}{\pi R_*^2}$ is the surface density of the star, $\Sigma$ 
is the surface density of the AGN disk and $T$ is the Keplerian period, { $R_{\rm R}= (m/3 M_\bullet)^{1/3} r$
is the Roche radius. With large relative speed between them, the accretion of disk gas onto stars with retrograde orbits
is negligible.  Also note that drag heating increases $R_\star$, $r_t$ and decreases $\Sigma_\star$ and $\tau_{\rm ret}$.
Since $\tau_{\rm ret} \ll \tau_{\rm pro}$, } retrograde stars primarily contribute to the in-disk TDE rate due to their shorter damping timescale \citep{Mckernan2022}.

Stars with initial inclination above the specific height $h$ of the AGN disk  
have their orbital inclinations damped by disk-star interactions, eventually being captured in prograde orbits. This effect prevents retrograde orbit stars from being refueled by disk capture \citep{Generozov2023, Wang2023b}. Thus, the total mass budget for retrograde in-disk TDEs is determined by the geometric intersection of the AGN disk and NSC: 
\begin{eqnarray}
M_{\rm ret}\sim \int_{r_t}^{r_m} dr \int_{-1}^{\cos(\pi-h)}\pi r^2 \rho_{\rm NSC}(r) d\cos I\,,
\label{eq:ret-budget}
\end{eqnarray}
yielding $M_{\rm ret} \sim \frac{2h(r_m)}{3}M_{\bullet}$ for a cluster with radial density power-law index $\gamma_{\rm NSC} = 2$.

\begin{figure}
    \includegraphics[width=\columnwidth]{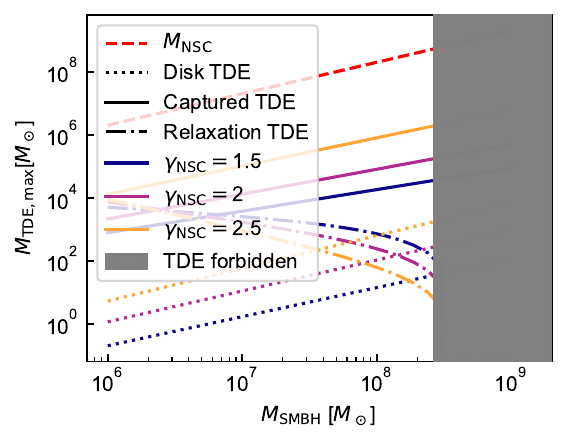}
    \caption{Total mass budget for disk/captured TDE and total disrupted mass for relaxation TDE during the lifetime (assumed to be $10^8$ yr) of the AGN for various SMBHs and NSCs. The dashed red line indicates the total mass of the NSC.}
    \label{fig:budget}
\end{figure}
The dotted lines in Figure~\ref{fig:budget} show the total mass budget of disk TDE for different masses of SMBH and NSC power-law index $\gamma_{\rm NSC}$. Despite their contributions, the total mass budget remains below $0.001\%$ of the NSC mass. 

Similar to relaxation TDE, we can define $\lambda_{\rm disk}(E)$ to calculate the rate of in-disk TDE
\begin{eqnarray}
\lambda_{\rm disk}(E) = T/\tau_{\rm ret}=\frac{\Sigma}{\Sigma_*}\frac{1}{h}\,.
\label{eq:disklossrate}
\end{eqnarray}
In the limit of small $h$, $\Sigma / 2 h r$ reduces to the midplane density, and $\lambda_{\rm disk}$ remains finite and small.

\subsection{Captured TDE from disk-star interactions}
In addition to stars initially situated in the AGN disk, stars from the NSC interact with the AGN disk due to aerodynamic drag. This interaction alters the stars' orbital properties, eventually resulting in their capture by the 
disk \citep{artymowicz1993, Generozov2023, Wang2023b}. This process preserves the quantity $L\cos^2(I/2)$ (where $L$ is specific angular momentum and $I$ is orbital inclination), as proven in \cite{Wang2023b}. The resulting captured semi-major axis can be given by
\begin{eqnarray}
a_{\rm f} = a_0(1-e^2_0)\cos^4(I_0/2)\,,
\end{eqnarray}
when the final orbit is circularized and coplanar. The captured objects will shrink their semi-major axis by a factor of $\frac{1}{(1-e_0^2)\cos^4(I_0/2)}$, where $a_0$, $e_0$ and $i_0$ are the initial semi-major axis, eccentricity and inclination, respectively.  
Stars with initial inclinations exceeding a critical value,
\begin{eqnarray}
I_{\rm TDE} &=& 2\arccos\left(\left(\frac{r_t}{a_0(1-e_0^2)}\right)^{1/4}\right)\nonumber\\
&\sim& \pi-2\left(\frac{r_t}{a_0(1-e_0^2)}\right)^{1/4}
\end{eqnarray}
will be tidally disrupted by the SMBH during the disk capture process.

\begin{figure*}
    \includegraphics[width=2\columnwidth]{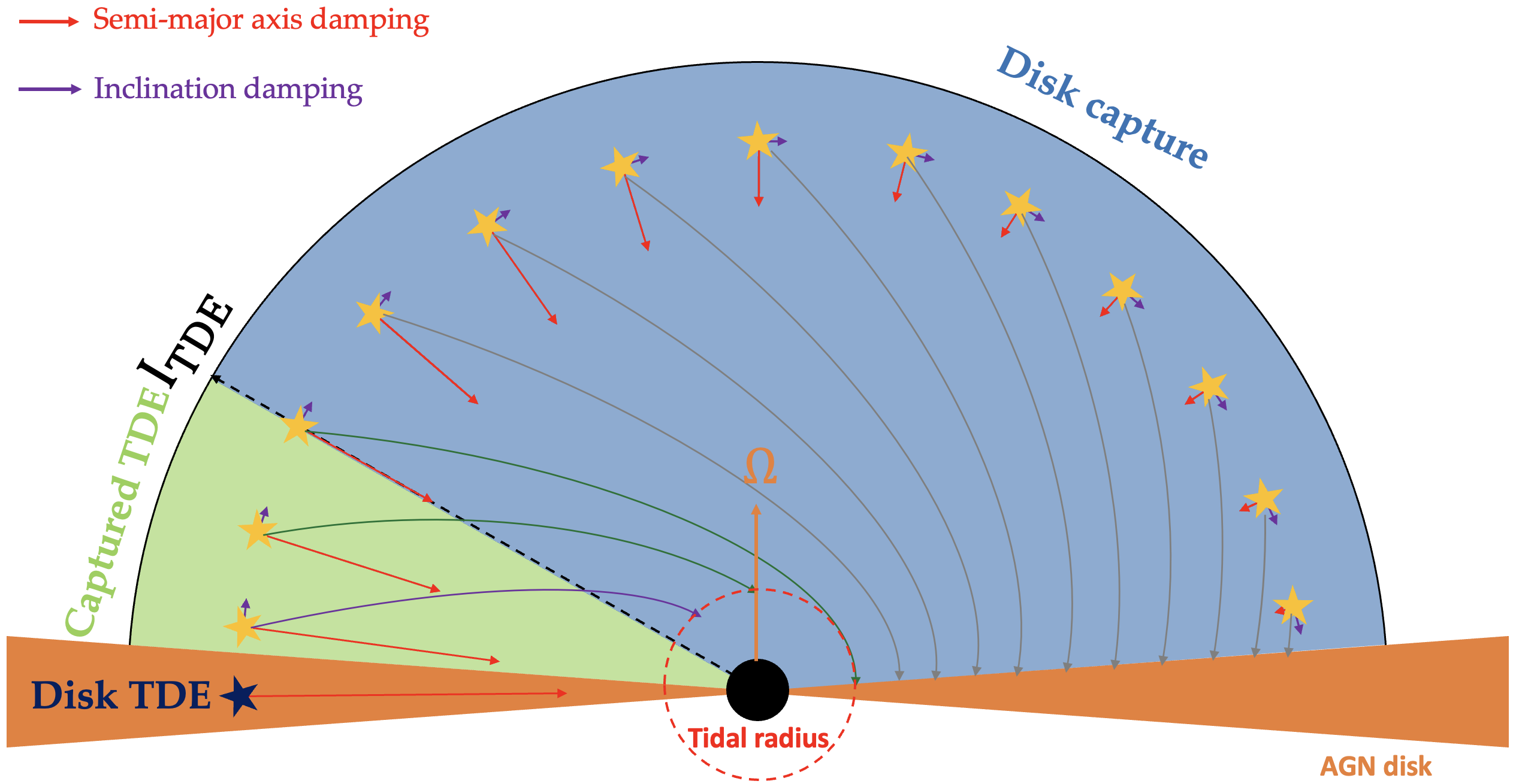}
    \caption{Schematics of captured TDE from disk-star interactions.}
    \label{fig:capture-TDE}
\end{figure*}
Figure~\ref{fig:capture-TDE}  illustrates this type of captured TDE, highlighting an additional population of NSC stars with high orbital inclinations prone to disruption during the AGN phase. The associated timescale for this disk-captured TDE is given by \citep{Wang2023b} 
\begin{eqnarray}
    \tau_{\rm cap}\sim \frac{\Sigma_*}{\Sigma}T { \sim {\tau_{\rm ret} \over h}}\,,
\end{eqnarray}
and the total mass budget can be estimated as
\small
\begin{eqnarray}
M_{\rm cap}\sim \int_{r_t}^{r_{\rm cap}} dr \int^{\rm arccos(I_{\rm TDE}(r))}_{\cos(\pi-h)}2\pi r^2 \rho_{\rm NSC}(r)d\cos I\,, \label{eq:cap-budget}
\end{eqnarray}
\normalsize
where $r_{\rm cap}$ is the truncation radius beyond which $I_{\rm TDE}$ is below the disk specific scale height.
Utilizing
\begin{eqnarray}
\lambda_{\rm cap}(E) = T/\tau_{\rm cap}=\frac{\Sigma}{\Sigma_*}
\end{eqnarray}
we can compute the rate of these captured TDE similar to relaxation TDE by substituting $\lambda (E)=\lambda_{\rm cap} (E)$ in 
Eq (\ref{eq:relaxlossrate}),
but with the total mass budget limited by Equation~\ref{eq:cap-budget}.
\begin{figure*}
    \includegraphics[width=\columnwidth]{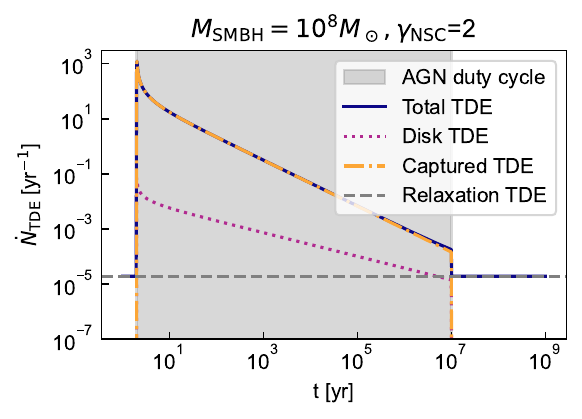}
    \includegraphics[width=\columnwidth]{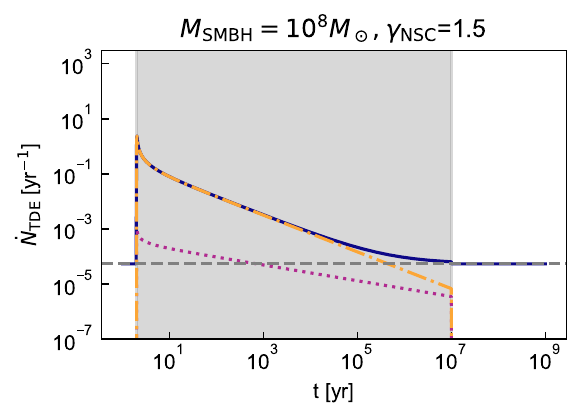}\\
    \includegraphics[width=\columnwidth]{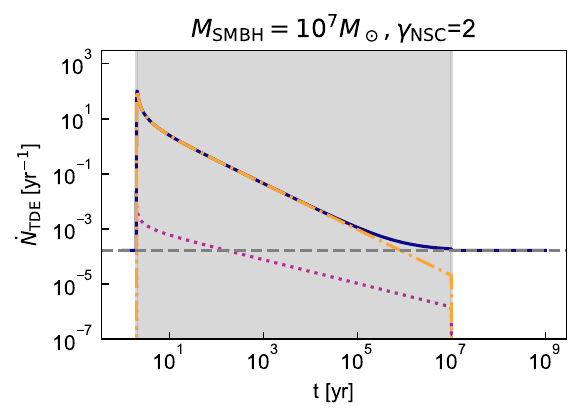}
    \includegraphics[width=\columnwidth]{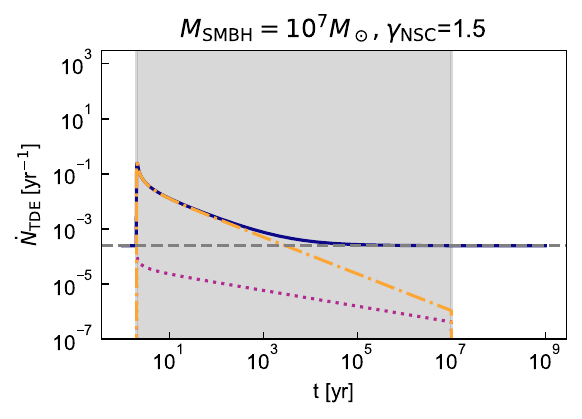}\\
    \caption{TDE rates of different mechanisms during the AGN phase for different SMBHs and NSC models.}
    \label{fig:rates}
\end{figure*}

Figure~\ref{fig:budget} and Figure~\ref{fig:rates} 
reveal the mass budget  (\ref{eq:ret-budget}, \ref{eq:cap-budget})
and TDE rates (Eq. \ref{eq:relaxlossrate} with appropriate $\lambda$ and $S_{\rm TDE}$), 
respectively, for different SMBH and NSC models. Captured TDEs 
 (with $\lambda (E)= \lambda_{\rm cap}$, $S_{\rm TDE}$: $r\in [r_t,r_{\rm cap}],\cos I\in[\cos(\pi-h),\rm arccos(I_{\rm TDE})], \phi \in [0, 2\pi]$) significantly outnumber disk TDEs 
(with $\lambda (E)= \lambda_{\rm disk}$, $S_{\rm TDE}$: $r\in [r_t,r_m],\cos I\in[-1,\cos(\pi-h)], \phi \in [0, 2\pi]$ ), owing to a much higher total mass budget, potentially constituting up to $1\%$ of the total NSC mass for steep NSCs. During the AGN phase, disk-star interactions predominantly enhance TDE rates through the disk capture mechanism, especially for AGNs hosting massive SMBHs ($\sim 10^8M_\odot$).

\section{CL-AGN activities from captured TDE}
The weak broad H$_\alpha$ emissions and Swift/XRT observations of X-ray emissions \citep{Parker2016, Mathur2018, WangJ2023} indicate that 
the ``turn-on" timescale of about 1 yr for some repeating CL-AGNs coincides with the typical TDE timescale \citep{Rees1988}, described by the equation:
\small
\begin{eqnarray}
\Delta t_{\rm TDE}\sim 0.35\left(\frac{M_\bullet}{10^7M_\odot}\right)^{1/2}\left(\frac{m}{1M_\odot}\right)^{-1}\left(\frac{R_*}{R_\odot}\right)^{3/2} \rm yr\,.
\end{eqnarray}
\normalsize
This similarity in timescales positions TDEs as potential mechanisms underlying CL-AGN behavior. The fraction of CL-AGNs relative to all AGNs can be approximated by\begin{eqnarray}
\frac{N_{\rm CL}}{N_{\rm AGN}}\sim \Delta t_{\rm TDE}\dot{N}_{\rm TDE}\,,
\end{eqnarray}
which results in a range of $10^{-5}-10^{-4}$ for relaxation TDEs, aligning with the findings reported in \citet{Yang2018} and \citet{Yu2020}. Recent spectroscopy analysis of MIR variability-selected SDSS partially obscured AGNs suggests a CL-AGN ratio as high as a few percent \citep{WangJ2023}, significantly exceeding earlier estimates. This elevated ratio challenges the viability of TDEs as an explanation, especially considering the relaxation TDE rate of a mere $10^{-4}-10^{-5}$ yr$^{-1}$ per galaxy. The occurrence of repeated CL activities on the timescale of decades years in some AGNs further complicates the viability of the TDE explanation.

Nevertheless, our findings illustrate a significant increase in the TDE rate by at least two orders of magnitude during the initial AGN phase due to disk-captured TDEs. This effect could be very significant in massive AGNs with a cuspy NSC. This enhanced rate $10^{-2}-10^{0}$ yr$^{-1}$ can plausibly account for the high incidence of repeating CL phenomena and the elevated CL-AGN ratio.

\begin{figure}
    \includegraphics[width=\columnwidth]{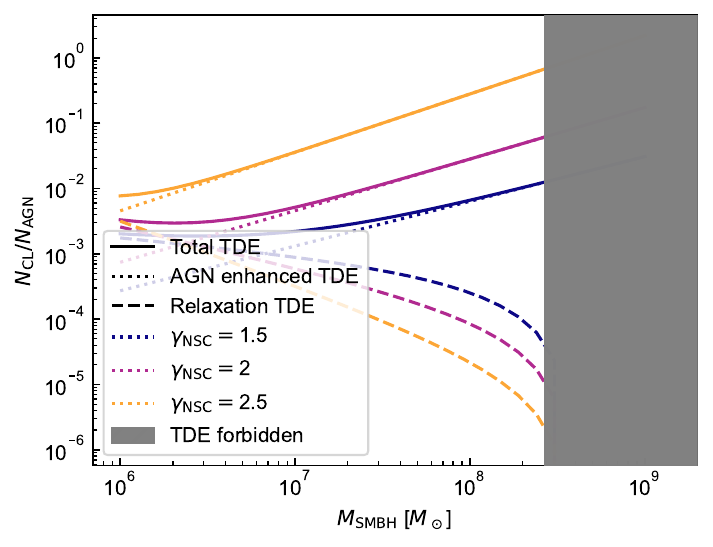}
    \caption{Fraction of CL-AGN among all AGNs as a function of SMBH masses for different TDE mechanisms.}
    \label{fig:ratio}
\end{figure}
Figure~\ref{fig:ratio} shows CL-AGN fractions amongst all AGNs, as anticipated by relaxation TDE and AGN-enhanced TDE for various SMBH masses and NSC models. Notably, the AGN-enhanced TDE rate predicts a CL-AGN fraction several orders of magnitude higher than that of the relaxation TDE rate, particularly for massive SMBHs accompanied by cuspy NSCs.

\section{Discussion and Conclusion}
\subsection{Relaxations with the presence of AGN disk}
The presence of the AGN disk gives rise to disk-star interactions. Over time, the disk will gradually capture stars from the NSC, which results in a depletion of stars during the relaxation process. Once the stars are captured by the disk, they tend to maintain a near-circular orbit due to the strong circularization induced by the disk-star corotation resonance for prograde stars and eccentricity damping from aerodynamic drag for retrograde stars. This makes it difficult for them to relax back into the loss cone region, characterized by small angular momentum. Consequently, the capture process may alter the relaxation process, rendering the relaxation TDE less efficient compared to scenarios with a dormant SMBH.

\citet{MacLeod2020} calculated the relaxation TDE rate in the presence of AGN disks. They discovered that the loss cone flux is only diminished in the inner region of the NSC, without significantly affecting the overall relaxation TDE rate. This is because the rate is predominantly determined by the full loss cone flux in the outer region.

\subsection{Captured TDE properties and connection to QPE}
For disk-captured TDEs, before the stars enter the tidal radius of the SMBH, the disk-star interaction continually damps the eccentricity of the stellar orbits. Consequently, as the star enters the tidal radius, its orbital eccentricity might have been significantly reduced by the disk capture process. Unlike in relaxation TDEs where stars approach the SMBH in a nearly parabolic orbit, stars in disk-captured TDEs will approach the SMBH with a much smaller eccentricity due to disk circularization. This means that disk-captured TDEs might be gentler than the typical TDEs from relaxations. Scenarios such as partial disruption or tidal peeling \citep{Ryu2020,Xin2023} might occur. {In this situation, the internal tidal dissipation could become sufficiently strong to cause runaway inflation for inspiraling stars on nearly circular orbits. As the star inflates, it is easier for the star to lose mass through the inner Lagrangian point which could slow down the inspiral rate \citep{gu2003}, extending the timescale for TDEs.}This might explain why some CL-AGNs with luminosity variations last longer than the timescale of normal TDEs.

More importantly, the light curve power-law index of $-5/3$ is predicted based on the frozen-in approximation with a fully disrupted star, wherein all disrupted materials fall into the SMBH on Keplerian orbits \citep{Rees1988, Lodato2009}. However, for partial TDEs and tidal peeling events, the existence of the remnant stellar core might alter the light curve power-law index from $-5/3$ \citep{Coughlin2019,Wang2021}. Therefore, the abnormal light curve power-law index in CL-AGNs cannot negate the potential TDE mechanism.

Since the disk-captured star could approach the SMBH in a nearly circular orbit, particularly if the circularization completes before the star enters the tidal radius, the interaction between the AGN disk and a closely circled stellar orbit may lead to quasi-periodic eruptions (QPEs) \citep{Miniutti2019,Giustini2020,Arcodia2021,Chakraborty2021, Linial2023}. These QPEs might also originate from the disk capture process, although more detailed hydro-simulations are necessary to confirm whether such a state can be achieved through disk-star interactions. If the majority of disk-captured TDEs enter this QPE phase, and disk-captured events are the underlying mechanism for CL-AGNs, QPEs could potentially serve as a precursor to CL-AGN "turn on".

\subsection{Conclusions}
In the analysis of tidal disruption events (TDEs) during the AGN phase of galaxies, we confirmed that beyond background relaxation TDEs, there exist two prominent mechanisms responsible for TDE occurrences: disk TDEs and captured TDEs. Disk TDEs emerge from the rapid semi-major axis damping experienced by fully disk-embedded retrograde stars, setting them apart from captured TDEs which occur when the disk apprehends stars from the NSC, provided the captured semi-major axis is lesser than the tidal radius of the supermassive black hole (SMBH).

In a comparative study of these mechanisms, it has been found that the rate of disk TDEs is considerably less than that of captured TDEs, especially noticeable around SMBHs with smaller masses. In the vicinity of low-mass SMBHs, the rate of disk TDEs can even be lower than the background relaxation TDE rate. Contrarily, captured TDEs boost the overall TDE rate significantly during the AGN phase, by several orders of magnitude than the background relaxation TDE rate, therefore indicating a potential mechanism for CL-AGNs.

Moreover, the disk-captured TDE mechanisms show unique signatures, distinguishing captured TDEs clearly from relaxation TDEs. The uniqueness of captured TDEs lies in the lower eccentricity of the stellar approaches to the SMBH, a consequence of the eccentricity damping experienced during the capture process. Unlike relaxation TDEs which see stars nearing the SMBH in nearly parabolic orbits, captured TDEs exhibit a more gentle approach, potentially resulting in phenomena such as partial TDEs, tidal peelings, or QPE precursors.

Consequently, the observable characteristics of captured TDEs are anticipated to vary significantly, possibly manifesting in divergent light curve slopes that could be the product of partial TDEs, tidal peelings, or QPE events. 


\section{Acknowledgments}
YW acknowledges useful discussions with Barry McKernan and Saavik Ford on CL-AGN observations and in-disk TDEs. YW, BZ and ZZ acknowledge NASA 80NSSC23M0104 and Nevada Center for Astrophysics for support. ZZ acknowledges support from the National Science Foundation under CAREER grant AST-1753168 and support from NASA award 80NSSC22K1413.

%

\vspace{5mm}







\bibliography{sample631}{}
\bibliographystyle{aasjournal}



\end{document}